\documentclass[12pt, titlepage]{article}

\usepackage{amsmath,amssymb}
\usepackage{graphicx}

\usepackage{hyperref, natbib, float}
\usepackage{lineno}

\usepackage[margin=1in]{geometry}

\usepackage{setspace} 
\author{\parbox{\linewidth}{\centering%
  Amy Willis, Department of Statistical Science, Cornell University  \endgraf\medskip
  John Bunge, Department of Statistical Science, Cornell University   \endgraf\medskip
   Thea Whitman, Department of Soil Science, University of Wisconsin-Madison }}
\date{ \vspace{1cm} Please direct correspondence to Amy Willis at \url{adw96@cornell.edu}}
\title{Improved detection of changes in species richness in high-diversity microbial communities}

\begin{document}
\maketitle

\section*{Abstract}
High throughput sequencing (HTS) continues to expand our understanding of microbial communities, despite insufficient sequencing depths to detect all rare taxa. These low abundance taxa are not accounted for in existing methods for detecting changes in species richness. We address this with a new hierarchical model that permits rigorous testing for both heterogeneity and biodiversity changes, and simultaneously improves Type I \& II error rates compared to existing methods.

\section{Introduction}
The diversity of a biological population is commonly used as a marker for ecosystem health \citep{globalwarming2, globalwarming1, karkman2011cold, lhkf, vaginalmicrobiome, deth}. However, species richness, a key element of diversity, is sensitive to changes in the ecosystem. Examples of affectors include temperature, time, biogeochemical conditions and  anthropogenic factors \citep{globalwarming2, globalwarming1, karkman2011cold, lhkf, vaginalmicrobiome, deth}. Understanding the mechanisms that may incite or accelerate changes in richness is crucial to sustaining ecosystem health. For this reason, many micro- and macroecologists are interested in formally testing for such changes in response to one or more covariates. 

In microbial ecology, while great gains have been made in reducing the cost and increasing the availability of  sequence data, only a small fraction of the population under study is usually sequenced.  As a result, in many sampled ecosystems, large components of biodiversity will be missing from the samples. While observed species richness (number of different species observed in the sample) is usually positively correlated with number of samples, if there are unequal sample sizes then the false conclusion of higher richness in the larger samples may be easily reached. Thus, in order to compare across samples of different sizes while focusing on the true total species richness (number of different species present in the population) in the ecosystem from which the sample was drawn, it is essential to consider the number of taxa missing from the samples, as well as the precision in predicting this number. The problem of substantial missing biodiversity is especially pronounced in many microbiomes, such as water, soil, gut and skin \citep{massana2015marine, file, deth, council2016diversity, grice2009topographical}.

Here we propose a method for modeling species richness that considers both the observed {\it and} unobserved members of the population. This allows us to draw conclusions about the population under study, rather than merely about the samples that were observed. This computationally efficient method has a number of key advantages. Most notably, it permits comparison across different sample sizes. Rigorous inference regarding the effects of covariates on biodiversity is possible. Multiple testing adjustments are implicit. Finally, it provides the first inferential method for assessing homogeneity of samples with respect to total biodiversity.

\section{Methods}

To achieve these goals, we propose a  three-component hierarchical  model for estimated species richness.  The first component  is an additive model, incorporating  covariate information  that is  known or believed  to influence true species richness. The second component captures the natural variability in species richness between different environments. The final component  is an error term  that accounts for the statistical error in estimating richness. We discuss the model, its estimation, and diagnostics below.

\subsection{Model details}
We wish to model the total species richness of $m$ populations. Denote the total richness in the $i$th population, observed and unobserved, by $C_i, i=1,\ldots,m$.  Also associated with each population is a set of $p$ covariates. We assume that species richness is a function of the covariates, but also a function of pure random variation, so that
\[
C_i = \beta_0 + \beta_1 x_{i,1} + \ldots + \beta_p x_{i,p} + u_i,
\]
where $x_{i,j}$ is the value of the $j$th covariate for the $i$th population, $\beta_j$ is its coefficient ($j=1,\ldots,p$), and $u_i$ is a random variable representing the variation in richness not attributable to the covariates.   We make the assumption that $u_1,\ldots,u_m$ are independent, identically distributed normal random variables\footnote{Independence is a very weak assumption in the HTS case because most samples will be sequenced distinctly. However, since a primary objective will be to test if $\sigma^2_u=0$, the assumption of normality should be verified, and we provide a diagnostic procedure in Section \ref{recs}. Identicality cannot be verified without replicates of every observed covariate combination, however if the other assumptions hold then the only consequence of non-identicality will be an inflated estimate of $\sigma^2_u$ and thus the tendency to conclude heterogeneity. In this way, the heterogeneity test can be considered to reflect both heterogeneity of richness {\it and} its variance: an interesting and unexpected consequence of this formulation.} with common variance $\sigma^2_u$. Additive nonlinear terms may be incorporated through the $x_{i,j}$'s as usual in a regression analysis.

Suppose the goal of the experiment is to investigate which covariates do and do not alter total species richness, or equivalently, which elements of $\beta = (\beta_1, \ldots, \beta_p)^T$ are equal to zero. In order to answer this question, we take a sample of individuals from each of the $m$ populations under study. We do not assume equal sample sizes or sampling depth, or that every taxon in each population was observed. Because we do not assume that every taxon in each population was observed, we do not know $C_i$ exactly for any $i$: the total species richness is unknown for each of the populations under study. Consequently our inference about the $\beta_j$'s requires accounting for error in estimation of the $C_i$'s. 

Based on each of our samples, we estimate $C_i$ by $\hat{C}_i$ with standard error $\hat{\sigma}_i$. A large number of estimators for species richness have been developed since the introduction of the problem into the statistical literature by \cite{fcwm} (see also \cite{bwwa} for a recent review). For estimators based on maximum likelihood or nonlinear regression, central limit theory ensures the asymptotic normality of estimates under the assumption of correct model specification \citep{rbmeunc}. We therefore make the broadly reasonable assumption that, conditional on the value of $C_i$, the estimate $\hat{C}_i$ is normally distributed around $C_i$ with standard deviation $\sigma_i$, that is,
\[
\hat{C}_i|C_i = C_i + \epsilon_i,
\]
where $\epsilon_i \sim \mathcal{N}(0,\sigma^2_i)$, $i=1,\ldots,m$.  Unconditionally we then have the final model 
\begin{equation} \label{mainmodel}
\hat{C}_i = \beta_0 + \beta_1 x_{i,1} + \ldots + \beta_p x_{i,p} + u_i + \epsilon_i = 
\beta_0 + x_i^T \beta + u_i + \epsilon_i,
\end{equation} 
where $x_i = (x_{i,1}, \ldots, x_{i,p})^T$. Since the only available information about $\sigma_i$ is the standard error $\hat{\sigma}_i$ we substitute the latter for the former and henceforth refer only to $\sigma_i$.  The results of our simulation studies (Sections \ref{size}, \ref{power} and \ref{homog}) suggest that this substitution is reasonable; however in Section \ref{homog} we propose an alternative approach that may improve robustness to this assumption.

It is important to note that the stochastic nature of the estimated total diversity arises both from $u = (u_1, \ldots, u_m)^T$and $\epsilon = (\epsilon_1, \ldots, \epsilon_m)^T$, that is, through the inherent random variation of the $C_i$ around $\beta_0 + x_i^T \beta$ and through the random variation of $\hat{C}_i$ around $C_i$.   Procedures that model the observed diversity $c_i$ as a linear function of the covariates effectively set $\hat{C}_i = c_i$ but treat $\sigma^2_i=0$, thus treating the sample as the population and unobserved diversity as null. As discussed previously, this can cause significant problems when sample sizes differ, because observed diversity correlating with sample size confounds the source of the elevated richness. Furthermore, modeling approaches based on relative frequencies cause problems when a small selection of the community greatly expands: lower relative frequencies of the rare taxa give the impression of lost diversity. However, they may not have been lost to the ecosystem (and the richness may be unchanged) but merely appear less frequently in samples due to the greater abundance of other taxa.

The conceptual difference between the above model and the most common existing method for evaluating richness changes is illustrated in Figure 1, where we observe biological replicates  (same day, unamended samples) from a soil field trial \citep{whitman}. Sample richness always underestimates total richness, but by a different factor for each sample that can be estimated from the observed rare species structure \citep{bwwa}. For this dataset, the ratio of estimated true richness to sample richness ranged between 1.35 and 1.95. Furthermore, the precision in the estimate varies widely by sample. Our model allows us to place more weight on samples whose richnesses are known with greater precision. In comparison, the most common existing method for assessing changes in richness is to fit a  linear regression model to the observed species richnesses, which presumes that all taxa were observed and that no sampling variability exists.

\begin{figure}
\begin{center}
\includegraphics[trim= 10mm 10 10 10]{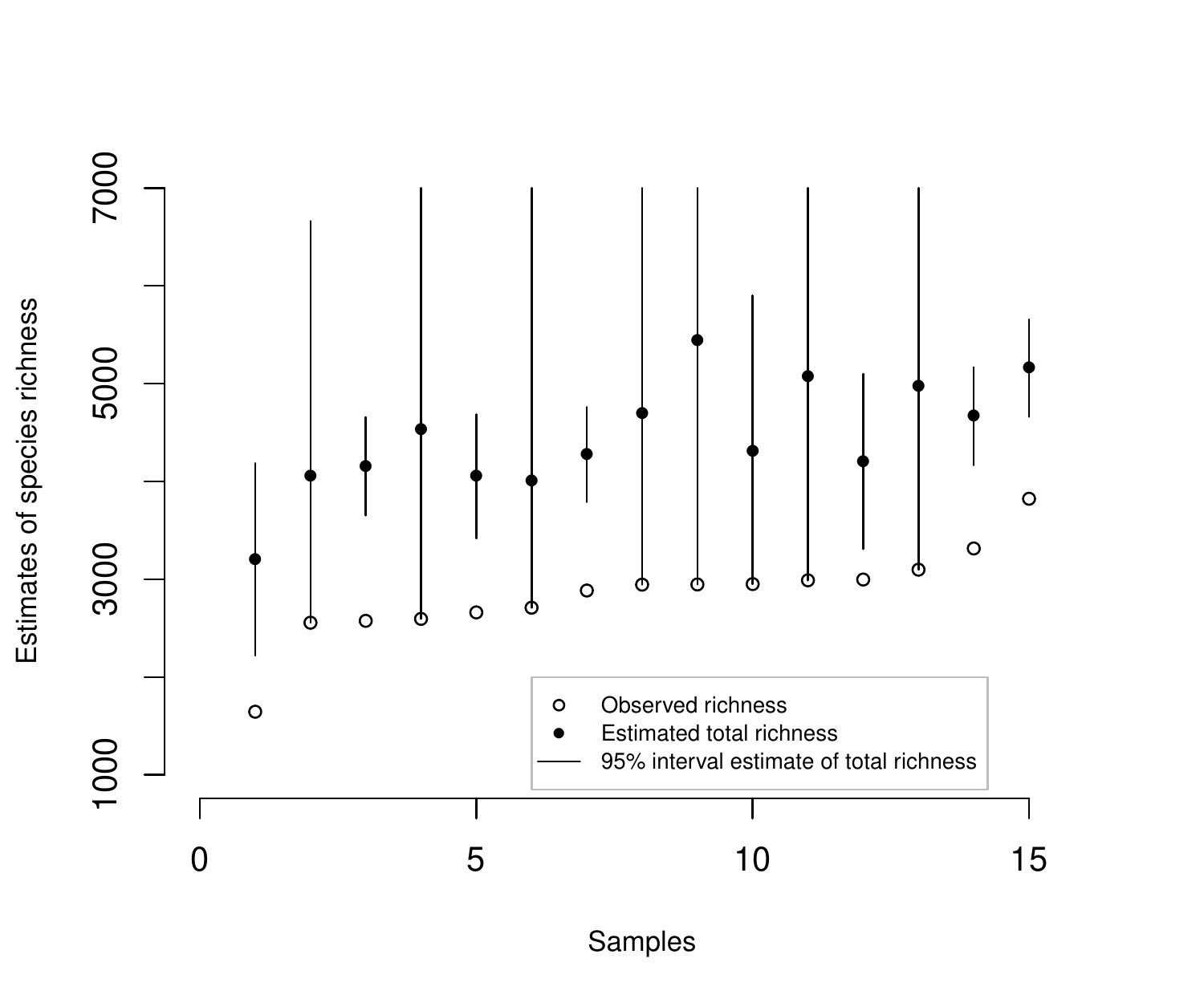}
\caption{Comparison of estimates of species richness for 15 biological replicates (no covariate effects) from the Whitman {\it et al.} dataset. Open symbols indicate observed richness and closed symbols indicate estimated total richness with 95\% confidence intervals. This ordering  of the replicates   suggests inconsistent levels of biodiversity across samples when only observed richness is considered. However, when total richness is considered along with its standard errors, homogeneity of richness is concluded ($p=0.169$). \label{illustrative_figure}}  
\end{center}
\end{figure}

Given model (\ref{mainmodel}) there are two main hypotheses of interest.  The first is $H_0: \sigma^2_u = 0$; that is, the variation in the true species richnesses across the $m$ populations is wholly attributable to the covariates $x_1, \ldots, x_p$ with no unexplained random variation. This hypothesis is often referred to as that of {\it homogeneity}. The alternative hypothesis of  {\it heterogeneity}, $H_A: \sigma^2_u>0$, supposes that there is more variability in the diversity estimates than can be explained by sampling-based variation in the estimates alone, and that some other mechanism (which we ascribe to the random variables $u_1, \ldots, u_m$) contributes to the observed discordance of the estimated species richnesses.  Possible interpretations of heterogeneity include model specification, missing predictors, or true biological heterogeneity between the ecosystems under study. While model misspecification may be diagnosed (see Section \ref{recs}), the distinction between the other options may only be informed by the scientific literature concerning the ecosystem under study.

The second main hypothesis of interest is $H_0: \beta_1 = \ldots = \beta_p = 0$, or alternatively, that none of covariates explains the variation in richness across populations.  The alternative hypothesis is then that at least one of the covariates affects richness.  If $H_0$ is rejected then interest focuses on the covariates that do influence richness: which $\beta_j$ are nonzero and what are their magnitudes? The relevant null hypothesis for the case of one variable is then $H_0: \beta_j=0$. Note that the usual regression interpretation of the coefficients applies and that $\beta_j$ is the expected increase in the true diversity of any of the $i$ populations for a one unit increase in $x_{i,j}$.

\subsection{Computation and optimization}

We now discuss estimation of the model parameters $\beta$ and $\sigma^2_u$, and implementation of the stated hypothesis tests. The log-likelihood of our model is $$l(\beta_0, \beta,\sigma^2_u | x_1,\ldots, x_m,\hat{C}_1,\ldots,\hat{C}_m, \sigma^2_1,\ldots,\sigma^2_m) = -\frac{1}{2} \sum_{i=1}^m \left[ \ln \left(\sigma^2_u + \sigma^2_i\right) + \frac{(\hat{C}_i-\beta_0 - x_i^T\beta)^2}{\sigma_u^2 + \sigma^2_i} \right].$$
Maximum likelihood (ML) is a natural choice of parameter estimation technique due to its many asymptotic and finite sample optimality properties in standard settings \citep{casellaberger,godambe}. However, in this application the choice to use ML is non-trivial because of the {\it boundary problem}: $\sigma_u^2 \geq 0$. This problem was studied by \cite{ccdr}, who demonstrate the failure of the usual likelihood ratio test asymptotics when testing $\sigma^2_u=0$ against $\sigma^2_u >0$.

Fortunately, we can exploit the well-developed literature on meta-analysis to resolve these difficulties. Meta-analyses arise in many social and health sciences where a researcher wishes to pool a number of different studies to determine the presence of an overall effect. Each richness estimate fulfills the role of a study's effect estimate, the standard error of the richness estimate fulfills the role of the standard error of the effect estimate, and the $m$ samples reflect $m$ different studies to be pooled. A comprehensive treatment of meta-analyses is given by \cite{demi}, who discusses both restricted maximum likelihood algorithms and also the best choice of hypothesis test in this nonstandard boundary case. We note also that in species richness comparison, as with meta-analyses, we only know the standard error in the estimates $\hat{\sigma}_i$ and not the true standard deviations $\sigma_i$. For this reason we base our choice of asymptotics on the results of \cite{demi} rather than those of \cite{ccdr}. Thus our restricted ML procedure maximizes 

$$l_R(\beta_0, \beta,\sigma^2_u ) = -\frac{1}{2} \left\{ \sum_{i=1}^m \left[ \ln \left(\sigma^2_u + \sigma^2_i\right) + \frac{(\hat{C}_i-\beta_0-x_i^T\beta)^2}{\sigma_u^2 + \sigma^2_i} \right] +\ln \left| \sum_{i=1}^m\frac{1+x_i^Tx_i}{\sigma^2_u+\sigma^2_j}\right| \right\},$$
and we denote the maximizing values by $\hat{\beta}_0$, $\hat{{\beta}}$ and $\hat{\sigma}^2_u$. Unfortunately there do not exist closed form expressions  for the estimates but we find that the range-restricted variable metric algorithm of \cite{byrd1995limited} in its implementation by \cite{rcore} is a fast and stable maximization algorithm for our restricted likelihood.  Our investigations suggest that the least squares estimates of $(\beta_0, \beta)$ obtained by regressing the covariates on the  richness estimates $\hat{C}_i$ are reasonable starting values for $(\hat{\beta}_0, \hat{\beta}),$ and the empirical variance in the estimates $\hat{C}_i$ is a reasonable starting value for $\hat{\sigma}^2$.

\subsection{Inference}
Because there are no boundary complications for $\beta$, its hypothesis testing falls in the standard Wald-type framework. Inverting second derivatives of the restricted log-likelihood gives the variance estimate

$$\hat{\text{Var}}(\hat{\beta}) = (\mathbf{X}^T \hat{\mathbf{W}}^{-1}\mathbf{X})^{-1},$$ 
where $\mathbf{X} = (x_1^T, \ldots, x_m^T)^T$ and $\hat{\mathbf{W}} = \text{diag} (\hat{\sigma}_1^2 + \hat{\sigma}_u^2,\ldots,\hat{\sigma}_m^2 + \hat{\sigma}_u^2),$
which we use to make marginal inference about the effect of each predictor on species richness via the test statistic $\frac{\hat{\beta}_i}{\sqrt{\hat{[\text{Var}}(\hat{\beta})]_{ii}}}$, which is distributed approximately $\mathcal{N}(0,1)$. The global test of $H_0: \beta_1 = \ldots = \beta_p=0$ has test statistic

$$\hat{\beta}^T \mathbf{X}^T \hat{\mathbf{W}}^{-1}\mathbf{X}\hat{\beta},$$
which is distributed asymptotically according to a $\chi^2_p$ distribution. Finally, we define our $Q$-statistic as 

$$Q = \sum_{i=1}^m \frac{(\hat{C}_i- \hat{\beta}_0- x_i^T\hat{\beta})^2}{\hat{\sigma}_i^2}.$$ Under the null hypothesis of homogeneity, $Q$ follows a $\chi^2$ distribution with $m-p-1$ degrees of freedom.

\section{Results}  \label{eval}

The proposed procedure, which we name \url{betta}, presents advantages in both Type I and Type II error rates in comparison to regression procedures, which we observe under simulation. The simulation methods underpinning both the size and power estimates (which we also believe to be a novel contribution) were designed to reflect data structures observed in microbial settings, and thus are intended to be realistic estimates of the method's advantages. The method's performance under the trivial case of negative binomial simulation structures even exceeds that shown below, but due to its limited ability to reflect microbial data structures is not included here. Furthermore, we examine two important questions in microbial ecology using the procedure: heterogeneity of soil communities, and dynamics of microbial communities in the human gut in response to an antibiotic. In the first instance, our analysis provides new insights on the community in question, and in the second instance we apply our method to rigorously confirm the conclusions of the original study by \cite{deth}.  

\subsection{Size of covariate tests} \label{size}

In order to compare the Type I error rate (statistical {\it size}) for the covariate test under a realistic high-diversity data structure, we redraw samples according to the distribution of OTUs in actual microbial datasets. We set up the simulation in the following way: choose an observed OTU table, and for each OTU in that sample, ascribe it the cell probability according to its relative abundance in the sample. Then, to mimic the differing sample sizes (numbers of observed OTUs) that HTS generates, we randomly choose a sample size based on the distribution of sample sizes across all OTU tables in the study. We then draw this number of samples from a multinomial distribution with the cell probabilities described above, use these draws to construct a frequency count table, estimate the total richness and its standard deviation, and calculate the sample richness. We repeat this 20,000 times, each time choosing a different sample size, and drawing a new multinomial sample of this sample size. In this way, we mimic two key features of microbial datasets: the relative distribution of both rare and common OTUs, and the differing numbers of OTUs observed in successive samples.\footnote{Note that the actual population richness is irrelevant; resampling from an unchanging distribution is sufficient to evaluate the size of the test.}

Armed with realistic redraws reflecting microbial datasets, we partition the 20,000 sets of richness estimates into 2,000 samples of 10 replicates, and create a covariate unrelated to richness for which to test for falsely significant relationships. That is, for each replicate we randomly ascribe a value across the grid $\{10,20,\ldots,100\}$, such that every covariate is ascribed exactly one frequency count table. We then compare \url{betta}  modeling  a richness estimate suited to resampled data structure (\url{breakaway} for high-diversity cases and \url{CatchAll} for medium-diversity cases; see Section \ref{recs}) against a simple linear regression of the observed richness (denoted $c_i$ for sample $i$) on the covariate (the most common method in the literature for evaluating changes in richness \citep{globalwarming1, globalwarming3,lhkf,deth,rodentparasite,vaginalmicrobiome}). To compare the effect of more replicates, we also consider the partition of 1,000 samples of 20 replicates modeled across the grid $\{5, 10,\ldots,100\}$. 

The Type I error rates for resamples from the \cite{whitman} dataset (Sample S009) are shown in Table 1. For each partition and richness measure, we show the empirical error rate (proportion of null hypotheses determined significant) for levels of significance of 1\%, 5\% and 10\%, noting that we expect less than 1\%, 5\% and 10\% Type 1 error rates. For 5 out of 6 combinations, \url{betta} has lower Type I error rates compared to the regression procedures, with the error rates consistently halved. The improvement is greatest for less stringent thresholds: for a 10\% level of significance  \url{betta}  reduces the Type I error rate by a factor of 5.

The analogous table for the \cite{deth}  dataset (Sample A1) is shown in Table 1, though in accordance with the experimental design of this dataset, the covariate under assessment is categorical rather than continuous (for each partition, half of the samples are assigned Category A, and the other half category B). In this case, we see that \url{betta} remains either accurate or  conservative: it consistently maintains a Type I error rate equal to or less than claimed. The improvement is less pronounced than in the high-diversity case of Table 1, and ranges between a small loss in size and a small improvement in size when compared with the regression procedure. Repeating the simulation with the continuous and discrete covariates reversed for the two datasets suggests that the differences between the two datasets shown in Table 1 are due to the differing data structures (high versus medium latent diversity) and not to testing a continuous versus categorical covariate. We conclude overall improvements in the Type I error rate for our method, with the improvement most pronounced in the high-diversity case and more stringent levels of significance.

\begin{table}
\caption{\label{fig3_t_whitman} Empirical Type I error rates for significance levels of $\alpha=(0.01, 0.05, 0.10)$ for the true null hypothesis of $\beta_1=0$ for 20,000 homogeneous redraws from the \cite{whitman} dataset, and 8,000 homogeneous redraws from the \cite{deth} dataset, each partitioned into samples of 10 and 20 replicates. } 
\begin{center}
\begin{tabular}{ l | l | c | c } \hline       
Dataset & Model & $n=10$ &  $n=20$ \\ \hline
\cite{whitman} & \url{betta} (\url{breakaway}) &  (0.016, 0.028, 0.039) & (0.006, 0.008, 0.021) \\
\cite{whitman} & Regression on $c$ & (0.006, 0.048, 0.093) & (0.013, 0.051, 0.103) \\ \hline
\cite{deth} & \url{betta} (\url{CatchAll})& (0.017, 0.063, 0.098) & (0.021, 0.050, 0.088) \\
\cite{deth} & Regression on $c$ & (0.016, 0.058, 0.100) & (0.011, 0.051, 0.099) \\ \hline
\end{tabular}
\end{center} 
\end{table} 

Before proceeding to compare Type II error rates, we emphasize that an appropriate richness estimation method is essential to the performance of \url{betta}. Choosing an overly restrictive estimator in a high-diversity case (e.g. \url{CatchAll} or Chao1) leads to artificially small standard errors due to model misspecification. In this case,  \url{betta} has reduced ability to detect no relationship between a covariate and richness ($H_0$), because the variability in richness estimates is falsely deflated. In the same way, choosing a highly flexible model in a medium or low-diversity case leads to reduced ability to detect a true relationship between a covariate and richness ($H_A$), because an inflated measure of variability overwhelms the true richness differences. It is for this reason that the above analyses were only conducted using richness estimators appropriate to the data structure. The authors maintain transparency with respect to the poor performance of  \url{betta} when modeling inappropriate estimates of richness and encourage practitioners to utilize the recommendations in the literature and in Section \ref{recs} with respect to the richness estimate  appropriate to their data structure. 

\subsection{Power of covariate tests} \label{power}

In order to  examine the ability of \url{betta} to detect true changes in richness (statistical {\it power}), we must introduce a richness gradient into the model. To maintain the realistic data structure of the size simulations, and to reflect that by its nature biodiversity loss almost always affects rare species \citep{chapin2000consequences}, we introduce this gradient along the rare species. 

For the \cite{whitman} dataset examined in the previous section, we create 1\% more multinomial categories and assign them each the same relative weight as the OTUs observed as singletons in the original data set (effectively creating 1\% more rare species). We ascribe this sample the covariate $x_{1,1}=1$. We repeat this for 2\% more multinomial categories for $x_{1,2}=2$, and so forth up to 20\% more multinomial categories with $x_{1,20}=20$. We repeat this 1,000 times to have 1,000 datasets with richness gradients and thus 1,000 p-values for which to assess the power of the test that $\beta_1=0$, and repeat with halving the partition to have 2,000 datasets with $n=10$ each. The results, shown in Table  2, show that \url{betta} is capable of enormous improvements in Type II error rate (the complement of power), the advantage being the most pronounced when the sample size is small and the desired Type I error rate is low. The improvement ranges between 1.8-fold to 18-fold improvement in power.

\begin{table}
\caption{\label{fig4_t_whitman}  Empirical power (rate of correct $H_A$ detection) at significance levels of $\alpha=(0.01, 0.05, 0.10) $ for the false null hypothesis of $\beta_1=0$ for inhomogeneous resamples mimicking the cell probabilities of the \cite{whitman} dataset, with a continuous richness gradient introduced. The same is shown for the \cite{deth} dataset with a 10\% increase in number of species.} 
\begin{center}
\begin{tabular}{ l | l | c | c } \hline       
Dataset & Model & $n=10$  &  $n=20$ \\ \hline
\cite{whitman} &  \url{betta} (\url{breakaway})  & (0.210, 0.295, 0.362) & (0.303, 0.502, 0.594) \\
\cite{whitman} & Regression on $c$  & (0.012, 0.076, 0.133) & (0.092, 0.225, 0.335) \\ \hline
\cite{deth} & \url{betta} (\url{CatchAll}) & (0.926, 0.950, 0.959) & (0.958, 0.986, 0.990) \\
\cite{deth} & Regression on $c$  & (0.044, 0.143, 0.230) & (0.094, 0.265, 0.390) \\ \hline
\end{tabular}
\end{center}
\end{table}

To evaluate the power in the medium-diversity case, and with a categorical covariate, we resample 20,000 datasets with 10\% more rare multinomial categories than the \cite{deth} dataset resampled in Section \ref{size}, and model the richnesses of 10 of the original resamples and 10 of the higher richness resamples, investigating the significance of the difference according to \url{betta} and to a regression on the observed richness. We note that even for this small increase in richness \url{betta} is extremely powerful, with greater than 92\% power to detect the change in richness. In comparison, regression methods never exceed 39\% power under this sampling scenario.

\subsection{Size and power of the homogeneity test} \label{homog}

We now turn our attention to evaluating the homogeneity test with respect to size and power\footnote{Note that no method for homogeneity determination of true species richness exists in the literature, and thus comparisons similar to Sections \ref{size} and \ref{power} are not possible for this test. As a result, we focus only on the Type I and II error rates.}. For the same size and power resamples from Sections \ref{size} and \ref{power}, we evaluate the Type I and Type II error rates for the null hypothesis that the samples are homogeneous with respect to richness in Table 3. We observe that for the high-diversity dataset of \cite{whitman}, the Type I error rate of the test is conservative, that is, for an $\alpha$-level test we observe less than an $\alpha$ rate of error. Furthermore, for this data structure the power is very high: 71\% for $\alpha=0.01$ with 10 data points and 99\% for 20 data points. For the medium-diversity datasets of \cite{deth}, the power is even higher: 77\% for $\alpha=0.01$ with 10 data points. However, this is at the expense of size, with Type I error rates up to double than controlled for. The explanation for this arises not from the richness comparison method, but from the richness estimation method. For the homogeneous replicates,  the average of the standard errors should match the standard deviation of the estimates (by definition). For the \cite{deth} dataset, the mean absolute deviation  of the estimates was 64.34 while the median of the standard errors was 58.20. Thus in this case \url{CatchAll} understates the true variability of its richness estimates, leading to inflated confidence and thus inflated risk of false determination of non-homogeneity. By comparison, for the \cite{whitman} dataset, \url{breakaway} overstates its standard errors, with a median absolute deviation of 126.24 but a median standard error determination of 189.10. This comparison provides a full explanation for the tendency of the method to favor $H_0$ (homogeneity) under \url{breakaway} and to favor $H_A$ under \url{CatchAll}. We advise practitioners to consider the results of the simulations carefully in conjunction with their own analysis of homogeneity determinations in the medium-diversity case. In particular, in order to maintain a Type I error rate of 7.4\%, we recommend rejecting at $\alpha=0.01$ when approximately 10 data points were obtained. Alternatively, the practitioner could resample from one of their own data structures in the same way as was performed above to find the appropriate $\alpha$ level for their number of samples and desired Type I error rate. The \url{R} code used to generate Table 3 is available from the corresponding author to facilitate such investigations.  

\begin{table}
\caption{ Empirical size and power of the homogeneity test of \url{betta}. The size estimates were derived from the same sampling scheme as in Section \ref{size}, and the power estimates as in Section \ref{power}. The homogeneity test captures the effect of absent predictors, thus no covariates were employed. Each cell shows results for significance levels of $\alpha = (0.01, 0.05, 0.10)$. \label{hom_t}} 
\begin{center}
\begin{tabular}{ l | c | c } \hline       
Dataset & $n=10$  &  $n=20$ \\ \hline
Size: \cite{whitman}  & (0.005, 0.006, 0.007) & (0.003, 0.003, 0.003) \\ 
Size: \cite{deth}  & (0.074, 0.136, 0.184) &   (0.126, 0.183, 0.244) \\ \hline
Power: \cite{whitman}  & (0.709, 0.735, 0.755) & (0.990, 0.994, 0.996)  \\ 
Power: \cite{deth}  & (0.774, 0.895, 0.935) & (0.968, 0.993, 0.997) \\ \hline
 \end{tabular}
 \end{center}
\end{table}

Note that when replicates are available, standard errors may be confirmed empirically. However, this may not be practical nor even possible under some experimental designs. In these cases, if the practitioner is skeptical of the standard errors produced by the richness procedure, a parametric bootstrap  approach can provide some information on its plausibility. Resampling from a multinomial distribution with cell probabilities equal to the sample's empirical taxa weights and passing the generated frequency count tables to the same estimator should give a collection of estimates whose standard deviation is close to the standard error of the original sample's richness estimate. Bootstrap theory gives us that this procedure should underestimate the standard error: a larger standard deviation of this collection compared to the standard error of the original sample suggests that errors are underestimated, and that the hypothesis of heterogeneity may be favored. The standard deviation of the resamples may be substituted for the standard error to correct for this.

\subsection{Application to homogeneity of soil communities}

To illustrate our test for species richness homogeneity of a highly diverse microbial environment, we investigate biological replicates from a soil field trial. Soil microbial communities are perhaps the most species-rich of all studied environments on Earth \citep{file}. Housing complex interfaces between the hydrosphere, atmosphere, lithosphere, and biosphere, soils exhibit extreme microscale heterogeneity in potential microbial habitats \citep{nwycr, trgk}, which may support the persistence of microbial species diversity \citep{lokn}. The complexity of these communities poses considerable challenges for diversity analysis and thus provide an interesting test case for the homogeneity hypothesis. 

\cite{whitman} extracted, amplified, and sequenced (Illumina MiSeq) bacterial 16S DNA with soils from a field trial with no amendments, with pyrogenic organic matter additions, and with fresh biomass additions. For each Day 1 sample with no amendment, \url{breakaway} \citep{breakaway}  was used to estimate the total  microbial OTU richness  in the soil due to the high-diversity nature of the data (singleton dominance). Confidence intervals for the estimates may be seen in Figure 1.  The \url{breakaway} algorithm failed to converge for one sample (Sample S026), which was thus excluded from the analysis. 

Because no (measured) covariates characterize differences between the samples, we fit an intercept-only model for estimates of species richnesses. Our method fails to reject the hypothesis that species richness is homogeneous between samples ($p=0.169$). Note that if only observed richness is considered, the samples appear to have different richnesses (Figure 1). Thus by accounting for the high variability across the samples we note that the samples are not distinct after the taxa that eluded detection are considered.

\subsection{Application to richness changes in the human gut}

We now demonstrate the method'€™s applicability in determining the effect of antibiotics on gut microbiome richness. \cite{deth} employed pyrosequencing to obtain rRNA sequences from the guts of three human subjects before, during, and after a course of ciprofloxacin. They observed that the treatment led to an overall decrease in the observed richness of the microbiota communities but were unable to test this formally.  \url{CatchAll} \citep{catchall}  was used to estimate the total microbial OTU richness due to the doubleton and tripleton dominance. No outliers were excluded. We fit our model for richness estimates with fixed treatment (pre-treatment, during treatment and post-treatment) and random patient effects (Figure 2) using the implementation \url{betta\_random}. We conclude that treatment is highly significant in decreasing richness ($p=0.027$), reducing richness by 494 species on average. However, we find that there is no significant post-treatment effect ($p=0.955$) and that richness recovers to pre-treatment levels after 4 weeks. This concurs with the visual conclusions of \cite{deth}, but we emphasize that this methodology provides a formal approach to testing their hypotheses. Note that within-patient heterogeneity of the gut microbiome \citep{davenport2014seasonal, lu2014spatial, wu2011linking} may be observed in Figure 2. 

\begin{figure}
\begin{center}
\includegraphics[trim= 10mm 10 10 10, scale=1]{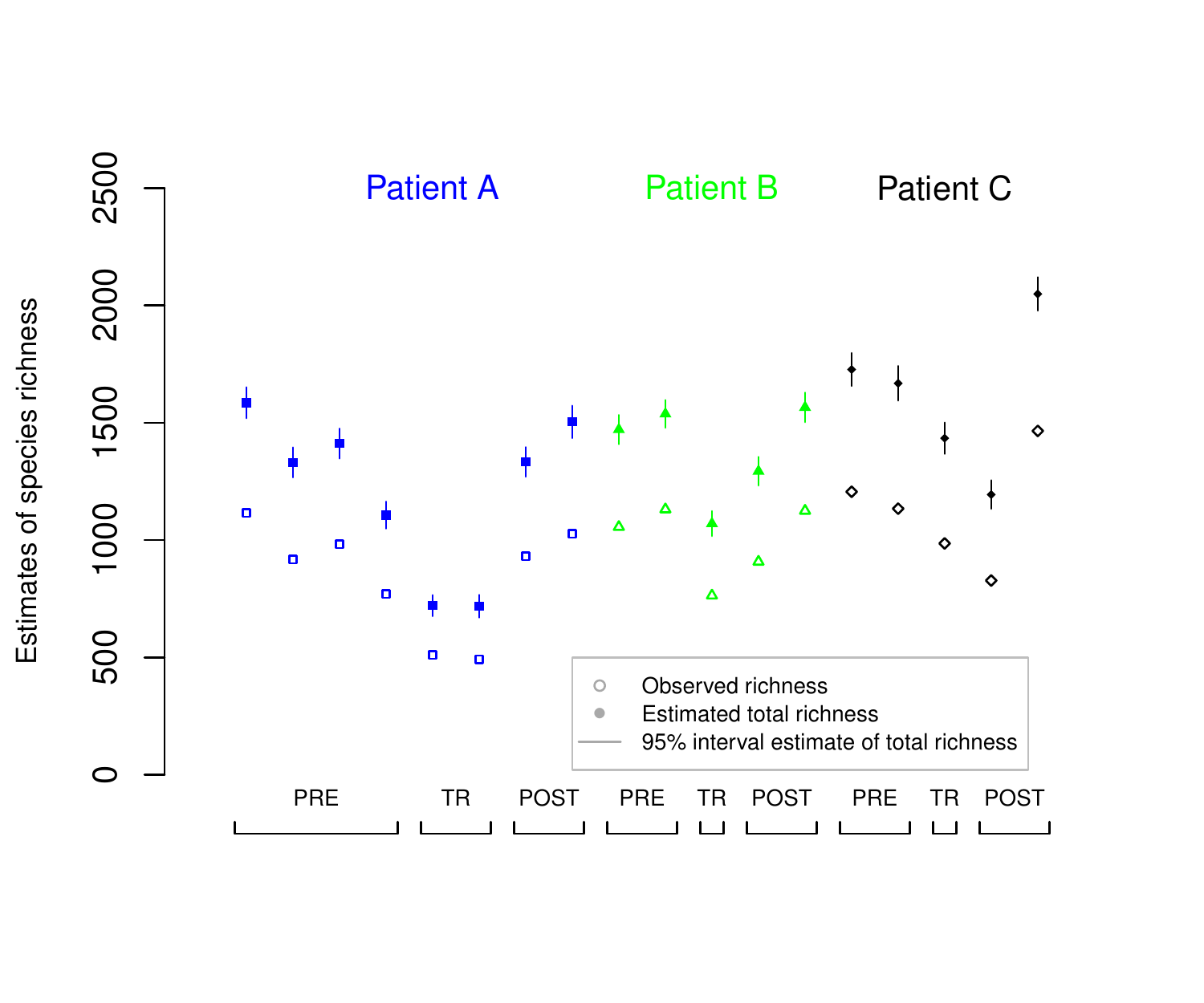}
\caption{Estimates of species richness of the gut microbiomes of three different human subjects before (PRE), during (TR) and after (POST) a course of ciprofloxacin  \citep{deth}. Each shape indicates a different patient, with open symbols indicating observed richness and closed symbols indicating estimated total richness. We conclude that the drug reduces gut richness by 494 taxa on average ($p=0.027$), however there are no longer significant differences in pre- and post- treatment levels  ($p=0.955$) after 4 weeks.    \label{illustrative_figure2}}  
\end{center}
\end{figure}

\section{Discussion}
We conclude that for high-diversity datasets, the greatest gain of the proposed methodology is with respect to its ability to correctly detect no change in richness, where the Type I error rate of the method is up to 5 times lower than the only currently available method. This is simultaneously achieved with power improvements, usually by a factor of around 3. Conversely, for medium-diversity datasets, the greatest gain is with respect to detection of true changes in richness, where for even small changes in richness the power can exceed 92\%.  Similarly, this is achieved with little to no loss in size. Both of these results are highly intuitive: large numbers of latent species destabilize richness estimates, thus increasing their variability and standard errors. When this variability is accounted for, supposed richness differences that are in fact small compared to the poor precision in the estimation are correctly detected as attributable to sampling variability. When variability in estimation is small, it is easier to detect changes, and the improvement in power is attributable to corrections for differing sample sizes across the covariate gradient. Overall it is clear that depending on the data structure either size or power can be greatly improved with almost no loss to the other. Furthermore, we have proposed a first formal test for homogeneity. It is particularly useful for comparing biological or technical replicates, and can be used to assess whether the experimental and computational procedures from sampling to final data output give consistent results. 

Nevertheless, due to the hierarchical nature of the model, a number of assumptions must be verified in order to produce valid results. The choice of species richness estimator, and perhaps more importantly the estimate of its standard deviation, is the most serious modeling choice in this method. Too much confidence in the precision of the richness estimate can induce false heterogeneity. Furthermore, over fitting, under fitting, or an inappropriate covariance structure across the samples can also compromise the validity of conclusions. Some practical guidelines are available below. We conclude with some comments of the potential applicability of the method to modeling other $\alpha$-diversity or community composition metrics, and statistical developments necessary for this generalization.

\subsection{Model selection and diagnostics} \label{recs}

The methodology is sensitive to the design matrix $\mathbf{X}$, and method of estimating $C$. Perusal of the richness estimates, and more importantly, their standard errors, is essential to ensure that the model is not overfit (with respect to predictors) and heterogeneity is not falsely concluded. One exploratory approach to diagnosing possible outliers and points of influence is to plot the estimated richness with error bars at $\pm 2$ standard errors. This technique derives from, and is limited by, the assumption that  estimated richnesses are normally distributed around the true  richness with standard deviation equal to the estimated standard error. The visual diagnostic for a point of influence is a tight interval (small estimated error), especially one centered far from the overall mean. While it is tempting to notice the large intervals in this type of plot, in fact these types of points do not exert a large influence on the model because their variability is captured in the large local error $\sigma_i^2$ rather than affecting the estimate of the global error $\sigma_u^2$. 


While some guidelines for appropriate richness estimators are available \cite{bwwa}, the authors consistently find that for high-diversity settings (the authors propose singleton-to-doubleton ratios of 1.5 or higher), the \url{breakaway} estimator \citep{rbmeunc} functions best with respect to plausibility of both estimates and standard errors. In medium- and low-diversity settings (singleton-to-doubleton ratios of 1.5 or lower), \url{CatchAll} \citep{catchall} functions well due to its stability, though as discussed below, its standard errors appear conservative in some medium-diversity settings. Note that in microbiome studies, the extent to which rare reads are discarded heavily affects richness estimates, and robustness of results to the quality control parameters should be thoroughly investigated. 

Finally, one assumption of our model is the normality of the $u_i$'s. Under this assumption, the distribution of $\left\{ \frac{\hat{C}_i-\hat{\beta}_0 - x_i^T\hat{\beta}}{\hat{\sigma}_i} \right\}_{i=1}^n$ should be approximately normal. Thus perusal of histograms and qqplots of these values should display approximate symmetry and no large outliers. Note that inferential tests for normality generally have poor power, and thus visual diagnostics should also be utilized.

\subsection{Generalization to other diversity indices}
While somewhat outside of the scope of this paper, we wish to briefly discuss the generalization of this method to other diversity indices. Any one-dimensional summary statistic  which estimates a population parameter and provides an estimate of its standard deviation can be modeled using the methodology described above. The asymptotics of the tests remain valid provided the distribution of the estimate is approximately normal around the true parameter with standard deviation close to its standard error. The main difficulty with extending this type of analysis to evenness indices is that standard errors in the estimates are rarely available, and bootstrap errors understate true sampling variability. There is emerging statistical research on estimators and error estimates for evenness indices \citep{zhzh,zhan}, and the authors hope that as the literature develops further the same analysis will be possible for a broad variety of diversity measures of interest.

\subsection{Closing remarks}
  
This method is the first inferential procedure for investigating homogeneity and response to covariates of species richness that focuses on the target of interest (population richness) rather than sample richness. This eliminates issues arising from different sample sizes or sampling depths, because precision is already reflected in the standard errors of the richness estimates. Allowing broad comparisons across microbial communities sampled to different depths, this procedure is capable of demonstrating factors affecting biodiversity and illuminating the presence or absence of heterogeneity across different ecosystems and processing pipelines.

The methodology, called \url{betta}, is available from CRAN via the {\url R} package \url{breakaway}. The random effects implementation is called \url{betta\_random}. Sample workflows are available from the corresponding author. Inquiries and extension requests are welcomed and should be directed to the corresponding author.

\pagebreak
\bibliography{betta}

\end{document}